\documentclass[a4paper,12pt]{article}

\usepackage{amsmath,amssymb}
\usepackage{graphicx}
\usepackage{color}
\newcommand{\strokedint}{\hspace{0.1cm}\rotatebox{15}{$-$}\hspace{-0.5cm}\int}
\newcommand{\nin}{\hspace{0.2cm}/\hspace{-0.4cm}\in}

\DeclareMathOperator{\Li}{Li}
\DeclareMathOperator{\sgn}{sgn}
\DeclareMathOperator{\Tr}{Tr}

\usepackage{geometry}
\numberwithin{equation}{section}
\allowdisplaybreaks

\begin{document}

\newcommand{\aff}[1]{${}^{#1}$}
\renewcommand{\thefootnote}{\fnsymbol{footnote}}

\begin{titlepage}
\begin{flushright}
{\footnotesize YITP-16-44}\\
{\footnotesize KIAS-P16060}
\end{flushright}
\begin{center}
{\Large\bf
Mass Deformed ABJM Theory on Three Sphere \\
in Large $N$ limit
}\\
\bigskip\bigskip
{\large Tomoki Nosaka\footnote{
\tt nosaka@yukawa.kyoto-u.ac.jp
}\aff{1},
Kazuma Shimizu\footnote{
\tt kazuma.shimizu@yukawa.kyoto-u.ac.jp
}\aff{2} and
Seiji Terashima\footnote{
\tt terasima@yukawa.kyoto-u.ac.jp
}\aff{2}
}\\
\bigskip\bigskip
\aff{1}: {\small
\it Korea Institute for Advanced Study, Seoul 02455, Korea
}\\
\aff{2}: {\small
\it Yukawa Institute for Theoretical Physics, Kyoto University, Kyoto 606-8502, Japan
}
\end{center}

\begin{abstract}
{
In this paper the free energy of the mass deformed ABJM theory on $S^3$ in the large $N$ limit is studied.
We find a new solution of the large $N$ saddle point equation which exists for an arbitrary value of the mass parameter, and compute the free energies for these solutions.
We also show that the solution corresponding to an asymptotically $\text{AdS}_4$ geometry is singular at a certain value of the mass parameter and does not exist over this critical value.
It is not clear what the gravity dual of the mass deformed ABJM theory on $S^3$ for the mass parameter larger than the critical value is.
}
\end{abstract}

\bigskip\bigskip\bigskip

\end{titlepage}

\renewcommand{\thefootnote}{\arabic{footnote}}
\setcounter{footnote}{0}

\tableofcontents

\section{Introduction}

The mass deformed ABJM theory \cite{HLLLP0,HLLLP,GRVV} is the theory obtained by deforming the three dimensional $U(N)_k\times U(N)_{-k}$ ${\cal N}=6$ superconformal Chern-Simons theory (called the ABJM theory) \cite{ABJM} with a set of relevant operators including mass terms for the bi-fundamental chiral multiplets.
While the ABJM theory describes the stack of $N$ M2-branes, the mass deformed ABJM theory is expected to describe the bound states of the M2-branes and the M5-branes through the fuzzy sphere configuration given in \cite{T,GRVV}.
This theory has special features which make it worth studying.
One of them is that the theory has the ${\cal N}=6$ supersymmetry, which is the (almost) maximum amount of the supersymmetries in three dimension.\footnote{
The mass deformations preserving fewer supersymmetries are also constructed in \cite{HLLLP,KKT}.
}
Nevertheless this theory is not conformal, hence has non-trivial dynamics and a renormalization group flow.
Furthermore, in the large $N$ limit this theory will have a gravity dual which should be obtained by a deformation to the gravity dual of the ABJM theory corresponding to the mass terms.
Therefore, this theory will be one of the basic models to be investigated in the large $N$ limit.

To study a supersymmetric field theory, we can use the localization technique \cite{W,Ne,P} which enables us to obtain the exact partition function as well as some supersymmetric correlators.
Each of these results is, however, given typically by a matrix model, i.e. an integration over the $N \times N$ matrix variables.
It is highly non-trivial to take the large $N$ limit in these matrix models.

In this paper, as in our previous work \cite{NST}, 
we continue to study the partition function $Z$ of the mass deformed ABJM theory on $S^3$ in the large $N$ limit.\footnote{
There are several large $N$ results for the mass deformed ABJM theory \cite{AZ,AR,NST}.
In \cite{AZ,AR} the authors analyzed the theory by continuing the Chern-Simons levels $k$ and $-k$ to complex numbers, and obtained the saddle point solution which is different from our solutions discussed in the following sections.
The solution in \cite{AR} may correspond to those discussed in appendix \ref{sec_anothersol}.
Also, in \cite{NST} we found two solutions in the region of small mass parameter $\zeta/k<1/4$.
We argue that one of them 
does not satisfy the saddle point equation at a boundary.
} 
We find a new solution of the large $N$ saddle point equation with an
arbitrary mass parameter and compute the free energy $F \sim N^2$ for
the solution.\footnote{
We call $F=-\log Z$ as the free energy even though we consider the theory on $S^3$.}
We also generalize the ansatz to obtain the free energy $F\sim N^{3/2}$ \cite{NST} in full extent, and find that the saddle point solution can not exist for the mass parameter larger than a certain critical value.
Because
the classical supergravity on an asymptotically $\text{AdS}_4$ spacetime
has $F\sim N^{3/2}$,
there would be no gravity duals for
the mass deformed ABJM theory on $S^3$ with the mass parameter larger
than the critical value.

This result seems surprising, as the critical mass is reached by a finite and relevant deformation from the ABJM theory.
Nevertheless, we can argue that this phase transition indeed occurs.
If the dimensionless mass parameter $m$, which is the mass parameter normalized by the radius of $S^3$, is small enough, the free energy $F$ will behave as $F\sim N^{\frac{3}{2}}$ since the theory reduces to the ABJM theory in the limit $m\rightarrow 0$.
The factor $N^{\frac{3}{2}}$ can be interpreted as $1/G_N$, hence this free energy is consistent with the classical supergravity.
On the other hand, if $m$ is sufficiently large we can integrate out the
bi-fundamental hypermultiplets first in the computation of the partition
function.
As a result we will obtain $F\sim N^2$.
Indeed, for the new solution we find the free energy scales like $F\sim N^2$ (see \eqref{freelargezr} and \eqref{free_finitez_new}).
Therefore, it is possible to have a phase transition in the
interpolating regime.\footnote{
The critical value of the mass parameter we found could be
different from this phase transition and represent another phase
transition, for which we do not have any physical reason to occur. 
}
The phase transition may be similar to the confinement/deconfinement transition if we regard the change of the mass parameter as a renormalization group flow.
We will discuss this aspect in \cite{ST}.

Note that this phase transition is absent in the ${\cal N}=2^*$ supersymmetric Yang-Mills theories on $S^4$ which is a four dimensional analogue of the mass deformed ABJM theory.
For this theory in the strong 't Hooft coupling limit
the saddle point solution and the free energy are smooth
under the change of the mass parameter \cite{RZ2,RZ1}.
Indeed, the free energy of the ${\cal N}=4$ supersymmetric Yang-Mills
theory, which is the massless limit of the theory, is $F\sim N^2$ also, 
thus both of the massless and the infinite mass limits 
are consistent with the gravity duals and
can be smoothly connected.

Needless to say, further investigations of the phase transition are desirable.
In particular, we should study the vacuum solution in the supergravity corresponding to the mass deformed ABJM theory on $S^3$ with an arbitrary mass parameter.
We also expect that this kind of phase transition will occur also in the other theories on $S^3$ describing the M2-branes in various backgrounds such as \cite{IK,HLLLP,TY1,ABJ}.
We hope to report on these in near future.

This paper is organized as follows.
In the next section we introduce the partition function of the mass deformed ABJM theory which is expressed as a $2N$ dimensional
integration.
We also write down the saddle point equations to evaluate the large $N$
limit of the partition function.
In section \ref{sec_finitek} and section \ref{sec_tHooft} 
we solve the
saddle point equations and determine the free energy $F=-\log Z$,
in the large $N$ limit for various values of the mass deformation
parameter. 
In section \ref{sec_finitek} we consider the problem in the limit $N\rightarrow \infty$ with $k$ kept finite.
In section \ref{sec_tHooft} we take the 't Hooft limit $k,N\rightarrow\infty$ with $k/N$ finite.
In both sections we also evaluate the vacuum expectation values of the $1/6$ BPS Wilson loops for the saddle point configurations and argue the interpretation of our results.
Section \ref{sec_disc} is devoted for discussion and comments on future directions.
In appendix \ref{sec_anothersol} we comment on another solution to the saddle point equations for finite $k$.
This solution give the free energy which is larger than that obtained in the same parameter regime in section \ref{sec_finitek}.
Appendix \ref{O1_finitez} contains the computation of the ${\cal
O}(1/N)$ corrections in the saddle point equations in section
\ref{sec_largez} and \ref{sec_anyk}, which are though irrelevant to the large
$N$ free energy.
In appendix \ref{newvarprob}, we rederive the solution which has 
the gravity dual in a similar way in \cite{NST}.

\section{Saddle point approximation of free energy}
\label{setup}
As in \cite{NST}, we will consider the mass deformed ABJM theory which
is the 3d ${\cal N}=6$ $U(N)_k \times U(N)_{-k}$ 
SUSY Chern-Simons matter theory with the Chern-Simons level $\pm k$ deformed by the mass terms and the interaction terms which preserve the ${\cal N}=6$ supersymmetry. 
The action of this theory on $S^3$ can be written as\footnote{
We take the radius of $S^3$ to be $r_{S^3}=1$ in this paper for notational simplicity.
}
\begin{align}
S_{\text{mABJM}}=S_{\text{ABJM}}+\frac{i\zeta}{2\pi}\int_{S^3} dx^3\sqrt{g}\bigl[\Tr(D-\sigma)+\Tr({\widetilde D}-{\widetilde\sigma})\bigr],
\end{align}
where $(\sigma,D$) are the auxiliary component fields in the $U(N)_k$ vector multiplet $(A_\mu,\sigma,\lambda_\alpha,D)$, and $({\widetilde \sigma},{\widetilde D})$ those in $U(N)_{-k}$ vector multiplet (see $\text{e.g.}$ eq(3.23) in \cite{HHL}).
Here $\zeta$ is a real parameter which is related to the mass of the matter fields as $m=r_{S^3}^{-1}\cdot \zeta/k$.

The supersymmetric gauge theories on the three sphere were studied in \cite{KWY,KWY2,J,HHL}, with the help of the localization technique.
For the mass deformed ABJM theory, it was found that the partition function is given by the following $2N$ dimensional integration
\begin{align}
Z=\prod_{i=1}^N\int d\lambda_id{\widetilde\lambda}_i
e^{-f(\lambda,{\widetilde\lambda})},
\label{ZN}
\end{align}
where
\begin{align}
f(\lambda,{\widetilde\lambda})&=
\pi ik\sum_{i=1}^N(\lambda_i^2-{\widetilde\lambda}_i^2)
-2\pi i\zeta\sum_{i=1}^N(\lambda_i+{\widetilde\lambda}_i)\nonumber \\
&\quad -\sum_{\substack{i,j=1\\(i>j)}}^N\log\sinh^2\pi(\lambda_i-\lambda_j)
-\sum_{\substack{i,j=1\\(i>j)}}^N\log\sinh^2\pi({\widetilde\lambda}_i-{\widetilde\lambda}_j)
+\sum_{i,j=1}^N\log\cosh^2\pi(\lambda_i-{\widetilde\lambda_j}).
\label{iff}
\end{align}
Here $\lambda_{i}$ and ${\widetilde\lambda}_i$ $(i=1,\ldots,N)$
respectively denote the eigenvalues of the scalar component field in the
vector multiplet for $U(N)_k$ and those for $U(N)_{-k}$, which are real
constant numbers characterizing the saddle point configurations of the
fields in the localization computation as
\begin{align}
\sigma=-D=
\begin{pmatrix}
\lambda_1&&&\\
&\lambda_2&&\\
&&\ddots&\\
&&&\lambda_N
\end{pmatrix},\quad
{\widetilde \sigma}=-{\widetilde D}=
\begin{pmatrix}
{\widetilde\lambda}_1&&&\\
&{\widetilde\lambda}_2&&\\
&&\ddots&\\
&&&{\widetilde\lambda}_N
\end{pmatrix},\quad
\text{(other fields)}=0.
\end{align}

In the limit of $N\rightarrow\infty$, these $2N$ integrations can be evaluated by using the saddle point approximation
\begin{align}
Z \approx e^{-f(\lambda,{\widetilde\lambda})},
\end{align}
with the eigenvalues $(\lambda,{\widetilde \lambda})$ being solutions to the following saddle point equations
\begin{align}
0&=\frac{\partial f(\lambda,{\widetilde\lambda})}{\partial
 \lambda_i}=2\pi ik\lambda_i-2\pi i \zeta-2\pi\sum_{\substack{j=1\\(j\neq i)}}^N\coth\pi(\lambda_i-\lambda_j)+2\pi\sum_{j=1}^N\tanh\pi(\lambda_i-{\widetilde\lambda}_j),\nonumber \\
0&=\frac{\partial f(\lambda,{\widetilde\lambda})}{\partial
 {\widetilde\lambda}_i}=-2\pi ik{\widetilde\lambda}_i-2\pi i \zeta-2\pi\sum_{\substack{j=1\\(j\neq i)}}^N\coth\pi({\widetilde\lambda}_i-{\widetilde\lambda}_j)-2\pi\sum_{j=1}^N\tanh\pi(\lambda_j-{\widetilde\lambda}_i).
\label{isaddle}
\end{align}
Note that $\lambda_i$ and ${\widetilde\lambda}_i$ can be 
complex numbers for the solutions to the saddle point equations, although the
original integration contour in the partition function \eqref{ZN} is the real axis.

For $\zeta\in\mathbb{R}$, as argued in \cite{NST}, 
we can consistently impose the following reality conditions to the eigenvalues:
\begin{align}
{\widetilde \lambda}_i=-\lambda_i^*.
\label{reality}
\end{align}
Under this assumption,
the saddle point equations \eqref{isaddle} reduce to
\begin{align}
&-ky_i-\sum_{\substack{j=1\\(\neq i)}}^N\frac{\sinh 2\pi(x_i-x_j)}{\cosh 2\pi(x_i-x_j)-\cos 2\pi(y_i-y_j)}\nonumber \\
&\quad\quad\quad\quad+\sum_{j=1}^N\frac{\sinh 2\pi(x_i+x_j)}{\cosh 2\pi(x_i+x_j)+\cos 2\pi(y_i-y_j)}=0,\label{isaddle_rr} \\
&kx_i-\zeta+\sum_{\substack{j=1\\(\neq i)}}^N\frac{\sin 2\pi(y_i-y_j)}{\cosh 2\pi(x_i-x_j)-\cos 2\pi(y_i-y_j)}\nonumber \\
&\quad\quad\quad\quad+\sum_{j=1}^N\frac{\sin 2\pi(y_i-y_j)}{\cosh 2\pi(x_i+x_j)+\cos 2\pi(y_i-y_j)}=0\label{isaddle_ri},
\end{align}
where $x_i$ and $y_i$ denote the real parts and the imaginary parts of the
eigenvalues
respectively, i.e.
\begin{align}
\lambda_i=x_i+iy_i.
\end{align}

In the following sections we will solve the saddle point equations \eqref{isaddle_rr} and \eqref{isaddle_ri}, and evaluate the free energy
\begin{align}
F=-\log Z \approx f(\lambda,{\widetilde \lambda}),
\label{iF}
\end{align}
for the solutions, which is written under the constraint \eqref{reality} as
\begin{align}
f(\lambda)&=-4\pi k\sum_{i=1}^Nx_iy_i+4\pi \zeta\sum_{i=1}^Ny_i-\sum_{\substack{i,j=1\\(i\neq j)}}^N\log\frac{\cosh 2\pi(x_i-x_j)-\cos 2\pi(y_i-y_j)}{2}\nonumber \\
&\quad +\sum_{i,j=1}^N\log\frac{\cosh 2\pi(x_i+x_j)+\cos 2\pi(y_i-y_j)}{2}.
\label{free_re}
\end{align}
We will also compute the vacuum expectation value of the supersymmetric Wilson loops
\begin{align}
W_R(C)&=\frac{1}{\dim R}\Tr_R \text{P}\exp\Bigl[\oint_C(iA_\mu dx^\mu+\sigma |dx|)\Bigr],\nonumber \\
{\widetilde W}_{\widetilde R}(C)&=\frac{1}{\dim {\widetilde R}}\Tr_{\widetilde R} \text{P}\exp\Bigl[\oint_C(i{\widetilde A}_\mu dx^\mu+{\widetilde \sigma} |dx|)\Bigr],
\end{align}
where $A_\mu$ and $\sigma$ are the component fields of the $U(N)_k$ vector multiplet and ${\widetilde A}_\mu$ and ${\widetilde \sigma}$ are those in the $U(N)_{-k}$ vector multiplet.
The closed path $C$ is an $S^1$ in $S^3$ which is determined by the 
supersymmetry used in the localization technique.
These Wilson loops preserves the $1/6$ of the ${\cal N}=6$ supersymmetry \cite{DPY,CW,RSY,KWY} and hence can be computed by the matrix model \eqref{setup} with the help of the localization method \cite{KWY}.
For simplicity we will consider only the Wilson loops with the fundamental representations, whose vacuum expectation values are given in the saddle point approximation as
\begin{align}
\langle W_\Box(C)\rangle&=\frac{1}{N}\sum_{i=1}^N\langle e^{2\pi\lambda_i}\rangle \approx \frac{1}{N}\sum_{i=1}^N e^{2\pi \lambda_i},\nonumber \\
\langle {\widetilde W}_\Box(C)\rangle&=\frac{1}{N}\sum_{i=1}^N\langle e^{2\pi{\widetilde\lambda}}\rangle \approx \frac{1}{N}\sum_{i=1}^N e^{2\pi {\widetilde \lambda}_i},
\label{WL}
\end{align}
with the substitution of the solution $(\lambda,{\widetilde\lambda})$ to the saddle point equations \eqref{isaddle}.\footnote{
Though the saddle point equations are modified with the insertion of the Wilson loops, the effects of such modifications are negligible for the fundamental representations.
}

Below we will assume $\zeta \geq 0$ without loss of generality; the results for $\zeta<0$ are easily generated with the help of the following $\mathbb{Z}_2$ ``symmetry'' of the partition function \eqref{ZN}
\begin{align}
\zeta \rightarrow -\zeta,\quad \lambda_i \rightarrow -\lambda_i,\quad \widetilde{\lambda}_i \rightarrow -\widetilde{\lambda}_i.
\end{align}
We will also denote $m \equiv \zeta/k$ which is the mass of the 
hypermultiplets.

\section{Large $N$ limit with finite $k$}
\label{sec_finitek}

In this section we study the saddle point equations for the free energy of the ABJM theory in the limit $N\rightarrow\infty$ with the Chern-Simons levels $k$ kept finite.

\subsection{Solutions in large $\zeta/k$ limit}
\label{sec_largez}

First, we consider the case $\zeta/k \gg 1$ (which is equivalent to the large radius limit of $S^3$ with a finite $\zeta/k$).
The saddle point equations further are simplified in this regime.
We take the following ansatz:
\begin{align}
\lambda_j=\frac{\zeta}{k}+i\frac{N}{k}+u_j+iv_j,
\label{largezshift}
\end{align}
where $u_j$ and $v_j$ are of ${\cal O}(N^0)$.
The shift in the real part $\zeta/k$ cancels the term $-2\pi i\zeta$, while the last terms in the saddle point equations \eqref{isaddle} are approximated as
\begin{align}
\sum_{j=1}^N\tanh\pi(\lambda_i-{\widetilde \lambda}_j)=N+{\cal
 O}(e^{-\frac{4\pi \zeta}{k}}),
\label{eq:1}
\end{align}
which is canceled by the shift in the imaginary part of the eigenvalues.
We are finally left with the following equations without $\zeta$
\begin{align}
-kv_i-\sum_{\substack{j=1\\(j\neq i)}}^N\frac{\sinh 2\pi(u_i-u_j)}{\cosh 2\pi(u_i-u_j)-\cos 2\pi(v_i-v_j)}&=0,\label{saddle_largez_r} \\
ku_i+\sum_{\substack{j=1\\(j\neq i)}}^N\frac{\sin 2\pi(v_i-v_j)}{\cosh 2\pi(u_i-u_j)-\cos 2\pi(v_i-v_j)}&=0.
\label{saddle_largez_i}
\end{align}
The free energy \eqref{free_re} also is simplified in this limit as
\begin{align}
f&=\frac{4\pi N^2\zeta}{k}+\delta f+{\cal O}(e^{-\frac{4\pi\zeta}{k}}),
\label{freelargezr}
\end{align}
with
\begin{align}
\delta f=-2N\log 2-4\pi k\sum_{i=1}^N u_iv_i-\sum_{\substack{i,j=1\\(i\neq j)}}^N\log \Biggl[2 \Bigl(\cosh 2\pi(u_i-u_j)-\cos 2\pi(v_i-v_j)\Bigr)\Biggr].
\label{deltaf}
\end{align}

Note that the equations \eqref{saddle_largez_r} and \eqref{saddle_largez_i} are in the same form as the saddle point equations of the matrix model for the Chern-Simons theory without the matter fields, which were analyzed in \cite{M,AKMV,HY,Su1} (with the pure imaginary Chern-Simons levels $k\rightarrow ik$).
In that sense the correction $\delta f$ in the free energy corresponds to the free energy of the pure Chern-Simons theory in the large $N$ limit.

\subsubsection{Eigenvalue distribution}

With the ansatz (\ref{largezshift}), the solution of the saddle point equations is the following:
\begin{align}
u_j=0+\frac{1}{N}g\Bigl(\frac{j}{N}\Bigr),\quad
v_j=\frac{j}{N}+n(j)+\Delta+{\cal O}\Bigl(\frac{1}{N}\Bigr).
\label{solI}
\end{align}
Here $g(s)$ is some function and $\Delta$ is a constant both of which being of ${\cal O}(N^0)$, while $n(j)$ is some integer which can be different for each $j$.
Indeed, after the substitution of these expressions the real part of the saddle point equation \eqref{saddle_largez_r} is of ${\cal O}(N^0)$, while the ${\cal O}(N)$ part of the imaginary part of the saddle point equations \eqref{saddle_largez_i} vanishes due to the following identity
\begin{align}
\sum_{\substack{j=1\\(j\neq i)}}^N\frac{\sin\frac{2\pi(i-j)}{N}}{1-\cos\frac{2\pi(i-j)}{N}}=0.
\label{period0}
\end{align}
Hence \eqref{solI} solves the saddle point equations up to ${\cal O}(N^0)$ corrections.

Let us evaluate the deviation of the free energy $\delta f$ for this solution.
The second term is obviously of ${\cal O}(N^0)$.
Approximating the cosine hyperbolic factor by $1$ we can compute the third term exactly as
\begin{align}
-\sum_{\substack{i,j=1\\(i\neq j)}}^N\log\Biggl[2\Bigl(\cosh 2\pi(u_i-u_j)-\cos 2\pi(v_i-v_j)\Bigr)\Biggr]\approx -N\log \prod_{i=1}^{N-1} 2\Bigl[1-\cos\frac{2\pi i}{N}\Bigr]=-2N\log N.
\end{align}
Hence the free energy in the large $N$ limit is
\begin{align}
f \approx \frac{4\pi N^2\zeta}{k} -2N\log N,
\label{flargez}
\end{align}
with the solution,
\begin{align}
\lambda_j \approx 
\frac{\zeta}{k}+i\Bigl(\frac{N}{k}+\frac{j}{N}-\frac{1}{2}\Bigr),
\label{solf}
\end{align}
where we have fixed the values of $\Delta$ and $n(j)$ as $\Delta=-\frac{1}{2}$ and $n(j)=0$, as discussed in appendix \ref{O1_finitez1}, though they actually do not affect the free energy \eqref{flargez}.

In the definition of the partition function, we neglected the 
$(1/N!)^2$ factor coming from the integration over $U(N)\times U(N)$.
Including this factor, the free energy becomes 
$f \approx \frac{4\pi N^2\zeta}{k}$.

There is an intuitive way of understanding our results above.
First recall that in the mass deformed ABJM theory the mass of the matter fields (adjoint hypermultiplets) is uniformly $m=\zeta/k$ which is induced by the Fayet-Illiopoulos term.
Hence in the regime $\zeta/k\gg 1$ the matter fields can be integrated separately as the massive free  hypermultiplets, which gives 
\begin{align}
Z_\text{hyper}(N)=\prod_{i,j=1}^N\frac{1}{(2\cosh\frac{2\pi\zeta}{k})^2}\approx e^{-\frac{4\pi \zeta N^2}{k}}.
\end{align}
This precisely reproduces the leading part of the free energy \eqref{freelargezr}.
On the other hand, after integrating out the matter multiplets in the
mass deformed ABJM theories we are left with the pure Chern-Simons theory (with the induced Yang-Mills terms).
The saddle point equations for the shifted eigenvalues $u_i+iv_i$ \eqref{saddle_largez_r} and \eqref{saddle_largez_i} can be interpreted as the saddle point equations for the partition function of this reduced theory.

Here we also comment on the F-theorem \cite{F1,F2}.
Our computations show that the free energy is an increasing function of 
mass parameter $m=\zeta/k$. 
However, at the IR fixed point the theory will be 
the $\mathcal{N}=2$ pure Chern-Simons theory which has smaller 
free energy than the one of the UV theory which is the ABJM theory.
Thus, our result is consistent with the F-theorem.
Indeed, in \cite{F2}, for free massive theory, the free energy 
was shown to be increasing function of the mass.\footnote{
Speaking more concretely, the leading part $4\pi N^2\zeta/k$ of the free energy \eqref{freelargezr} can be canceled by a local counter term $\Lambda\int_{S^3} dx^3\sqrt{g} (R+\cdots)$, as it is linear in the mass parameter $m=r_{S^3}^{-1}\cdot \zeta/k$.
Hence the F-theorem applies not to the whole free energy but only to $\delta f$ \eqref{deltaf}.
}

\subsubsection{Wilson loops}
\label{sec_<W>}

Here we shall compute the vacuum expectation values of the supersymmetric Wilson loops \eqref{WL}.
First consider the Wilson loop associated with $U(N)_k$ gauge group in $U(N)_k\times U(N)_{-k}$.
With the substitution of the saddle point configuration \eqref{largezshift} with \eqref{solI} we obtain
\begin{align}
\langle W_\Box(C)\rangle=\frac{1}{N}e^{\frac{\zeta}{k}+i\frac{N}{k}}\sum_{j=1}^N\exp\Bigl[\frac{2\pi ij}{N}+{\cal O}(N^{-1})\Bigr].
\label{WBoxcancel}
\end{align}
Similarly, the Wilson loop for $U(N)_{-k}$ can be computed as
\begin{align}
\langle {\widetilde W}_\Box(C)\rangle =\frac{1}{N}e^{-\frac{\zeta}{k}+i\frac{N}{k}}\sum_{j=1}^N\exp\Bigl[\frac{2\pi ij}{N}+{\cal O}(N^{-1})\Bigr].
\label{WtildeBoxcancel}
\end{align}
If we neglect the ${\cal O}(N^{-1})$ deviations in the exponent, the leading part of the right-hand side vanishes in both cases.
The vanishing of the leading part of the vacuum expectation values of the Wilson loops may have some physical implication, which will be discussed in \cite{ST}.

\subsection{Finite $\zeta/k$}
\label{sec_M}

Below we will consider the limit $N\rightarrow \infty$ with both $k$ and $\zeta$ kept finite.
In this limit the mass deformed ABJM theory is expected to correspond to
the eleven dimensional supergravity 
with some classical geometry which
will be asymptotically $AdS_4 \times S^7/Z_k$.

We first show that for any finite $\zeta/k$, there is a solution which
is a simple generalization of the solution obtained in the last section
and has the same expression for the free energy 
$f \sim 4 \pi N^2 \zeta/k$ in the large $N$ limit.
Next we study the solutions which has the free energies $f\sim N^{3/2}$.
We find that the solution to the saddle point equation is unique for $\zeta/k<1/4$.\footnote{
Note that this parameter regime was already analyzed in \cite{NST}, where we found the two solutions to the saddle point equations \eqref{isaddle_rr} and \eqref{isaddle_ri}.
As we will see later, however, we should impose the boundary conditions to the profile functions of the eigenvalue distribution (which were imposed by the minimization of the free energy against the continuous moduli of the solutions in the context of the previous studies \cite{HKPT,NST}).
One of the solutions in \cite{NST} is actually excluded due to these additional constraints.
} 
For $\zeta/k>1/4$, on the other hand, we find 
there are no solutions with $f \sim N^{3/2}$.

\subsubsection{Solution with $f\sim N^2$ for any $\zeta/k$} 
\label{sec_anyk}

Let us start with the small generalization of the ansatz in the last section \eqref{solf} ($\lambda_i=x_i+iy_i$)
\begin{align}
x_i=\frac{\zeta}{k}+\frac{1}{N}g\Bigl(\frac{i}{N}\Bigr),\quad
y_i=\frac{N}{k}+\frac{i}{N}+\Delta+\frac{1}{N}h\Bigl(\frac{i}{N}\Bigr),
\label{solnew}
\end{align}
with $g(s)$ and $h(s)$ some functions and $\Delta$ some real constant, both being of ${\cal O}(N^0)$.
Indeed we can show that the left-hand side of the imaginary part of the saddle point equations \eqref{isaddle_rr} vanishes with the help of the following trivial generalization of the identities \eqref{period0}
\begin{align}
\sum_{j=1}^N\frac{\sin\frac{2\pi(i-j)}{N}}{a+\cos\frac{2\pi(i-j)}{N}}=0,\quad\quad(a\nin (-1,1)).
\label{period02}
\end{align}
Similarly the ${\cal O}(N)$ terms in the real part of the saddle point equation \eqref{isaddle_ri} vanish due to
\begin{align}
\sum_{j=1}^N\frac{1}{\cosh b+\cos\frac{2\pi(i-j)}{N}}=\frac{N}{\sinh b}.\quad\quad(b>0)
\label{period1}
\end{align}
We can also solve the ${\cal O}(N^0)$ part of the saddle point equations to determine $(f(s),g(s),\Delta)$, though they are irrelevant to the leading part of the free energy.
The computation is parallel to those in the large $\zeta$ limit and displayed in appendix \ref{O1_finitez2}.

The free energy $f$ for this solution also takes the same form as in the case of the large $\zeta$ limit.
In the limit $N\rightarrow \infty$ the leading parts of the first two terms in \eqref{free_re} precisely cancel with each other, hence only the last two terms are relevant
\begin{align}
f(\lambda)&\approx -\sum_{\substack{i,j=1\\(i\neq j)}}^N\log\biggl[\frac{1-\cos\frac{2\pi(i-j)}{N}}{2}\biggr]+\sum_{i,j=1}^N\log\biggl[\frac{\cosh\frac{4\pi\zeta}{k}+\cos\frac{2\pi(i-j)}{N}}{2}\biggr]\nonumber \\
&=\frac{4\pi\zeta N^2}{k}+{\cal O}(N\log N).
\label{free_finitez_new}
\end{align}
To obtain the second line it is convenient to replace the summations over $i,j$ with the integrations of continuous variables $s\sim i/N$ and $s^\prime \sim j/N$ over $s,s^\prime\in (0,1)$.
The ${\cal O}(N\log N)$ denotes the error due to the difference between the integrations and the original discrete summation.

\subsubsection{Solutions with $f \sim N^{\frac{3}{2}}$}
\label{N^3/2gen}

Now we shall go on to the solutions with the free energy $f\sim N^{3/2}$.
We use the continuous notation $\lambda_i\rightarrow \lambda(s)$ with
$s\sim i/N +{\rm const.}$ 
and take the following form:
\begin{eqnarray}
 \lambda (s) &=& \sqrt{N} z_1(s)+z_2(s), \nonumber \\f
 \tilde{\lambda}(s) &=& \sqrt{N} z_1(s)-z_2(s),
\label{ag}
\end{eqnarray}
where $z_1$ and $z_2$ are $N$ independent 
arbitrary complex valued functions of $s$.\footnote{
The following generalization also gives the large $N$ scaling of the free energy $f\sim N^{3/2}$
\begin{align}
\lambda(s)=\sqrt{N}z_1(s)+z_2(s),\nonumber \\
{\widetilde\lambda}(s)=\sqrt{N}z_1(s)+z_3(s).
\end{align}
However, this ansatz is reduced to \eqref{ag} by an ${\cal O}(N^{-1/2})$-shift of $z_1(s)$ which is irrelevant to our leading analysis.
}

Note that the transformation
\begin{eqnarray}
 \tilde{\lambda}(s) \rightarrow  \tilde{\lambda}(-s),
\label{gauge}
\end{eqnarray}
only changes the ordering of the $U(N)$ index of the $\tilde{\lambda}$,
thus the gauge symmetry.
This means that the configuration $\{ \lambda(s), \tilde{\lambda}(s) \}$
is equivalent to  $\{ \lambda(s), \tilde{\lambda}(-s) \}$.
We can see that the form (\ref{ag}) includes 
the ansatz taken in 
\cite{NST} for pure imaginary $\zeta$ and 
for real $\zeta$ with the gauge transformation (\ref{gauge}).\footnote{
The large $N$ analysis in this section 
includes those in \cite{NST} and the simplest examples in \cite{HKPT,F1}.
Furthermore, as we will see below, 
the one in this section is much simpler than those.
}
Note that here we do not require the reality condition
(\ref{reality}).\footnote{
In the Appendix \ref{newvarprob},
we solve the saddle point equation imposing the reality condition,
which will be useful to compare the previous studies including \cite{NST}.
}

The above gauge symmetry also allows us to assume that $\mathrm{Re}(z_1(s))$ is a monotonically increasing function with respect to $s$.
For simplicity, in this section we shall further assume that the profile functions $z_1(s)$ and $z_2(s)$ are piecewise continuous in $0\le s\le 1$ for this choice of the ordering.

We believe that the form (\ref{ag}) is the most general form 
which gives $f \sim N^{\frac{3}{2}}$.
Of course, there are no proofs for this, however, 
there should be non-trivial cancellation of ${\cal O}(N^2)$ and 
${\cal O}(N^\frac{5}{2})$ terms in the free energy 
in order to obtain $f \sim N^{\frac{3}{2}}$, which makes 
finding other possible forms highly difficult.

We will evaluate the free energy for the configuration (\ref{ag}) which 
is indeed ${\cal O}(N^{\frac{3}{2}})$.
The Chern-Simons term, which is proportional to $k$,
and the FI term, which is proportional to $\zeta$, are 
easily evaluated to 
\begin{eqnarray}
 4 \pi N^{\frac{3}{2}} \int ds 
\left( i k \, z_1 \,  z_2 - i \zeta \, z_1
\right).
\end{eqnarray}
For other logarithmic terms in the free energy,
for example, 
\begin{eqnarray}
  N^2 \int d s' \int ds \ln \left(
\sinh^2 (\sqrt{N} \pi (z_1(s)-z_1(s') 
+\pi (z_2(s)-z_2(s') )\right) ,
\end{eqnarray}
we will use the decomposition
\begin{eqnarray}
 \int ds \ln (\sinh^2 (z)) &=&  
 2 \int ds  \sgn ({\rm R (s)}) \, z \nonumber \\
\,\,\, &&+  \int_{{\rm R (s) >0} } ds \ln (\sinh^2 (z) e^{-2z})  
+  \int_{{\rm R (s) <0} } ds \ln (\sinh^2 (z) e^{2z}),
\label{decom}  
\end{eqnarray}
where R($s$) is a real function,
and the decomposition which is obtained by replacing $\sinh$ by $\cosh$ in \eqref{decom}.
We take $\text{R}(s)={\rm Re} (z_1(s)-z_1(s'))$.
Then, we can see that the terms linear in $z$ cancel each others:
\begin{align}
   N^2 \pi \int d s' \int ds  
{\rm Re} (z_1(s)-z_1(s')) (
& -(\sqrt{N} (z_1(s)-z_1(s')) +z_2(s)-z_2(s'))\nonumber \\
& -(\sqrt{N} (z_1(s)-z_1(s')) -z_2(s)+z_2(s')) \nonumber \\
& +2 (\sqrt{N} (z_1(s)-z_1(s')) +z_2(s)+z_2(s') 
))=0.
\end{align}
Remaining terms can be evaluated by using a formula
(here dot $\cdot$ is the abbreviation for $\frac{d}{ds}$)
:
\begin{eqnarray}
 \int_{s_0} ds \ln (\cosh (z(s)) e^{-z(s)}) &\sim&
\frac{1}{\sqrt{N} \dot{u}(s)|_{s=s_0} }
\int_{C_+}
dt \ln (\cosh (t) e^{-t}), \label{logcoshapprox1} \\
 \int^{s_0} ds \ln (\cosh (z(s)) e^{z(s)}) &\sim&
\frac{1}{\sqrt{N} \dot{u}(s)|_{s=s_0} }
\int_{C_-}
dt \ln (\cosh (t) e^{-t}),
\label{logcoshapprox2}
\end{eqnarray}
for
$\dot{u}(s)|_{s=s_0}>0$
where
\begin{eqnarray}
 z(s)=\sqrt{N} u(s)+v(s),
\end{eqnarray}
$u(s_0)=0$ and the path $C_{\pm}$ is a straight line between 
$t=\pm v(s_0)$ and $t=\sqrt{N} \dot{u}(s)|_{s=s_0} $ with $N \rightarrow \infty$.
Note that the $\cosh$ in the formula can be replaced with $\sinh$.
Then, the remaining parts of the free energy is
\begin{eqnarray}
&& N^{\frac{3}{2}} \int ds' 
\frac{1}{\pi \dot{z}_1(s')} \left(
-4 \int_0^{\infty} dt \log (\sinh(t)e^{-t})
	  \right. \nonumber \\
&&\left. 
+2 \int_{2 \pi z_2(s')}^{\infty} dt \log (\cosh(t)e^{-t})
+2 \int_{-2 \pi z_2(s')}^{\infty} dt \log (\cosh(t)e^{-t})
\right) \\
&&= N^{\frac{3}{2}} \int ds' 
\frac{1}{\pi \dot{z}_1(s')} \left(
-4 \int_0^{\infty} dt \log (\frac{\sinh (t)}{\cosh (t)} )
+2 \int_{2 \pi z_2(s')}^{0} dt \log (\frac{\cosh(t)e^{-t}}{\cosh(t)e^t}) 
\right) \\ 
&&= N^{\frac{3}{2}} \int ds' 
\frac{1}{\pi \dot{z}_1(s')} \left(
\frac{1}{2} \pi^2 +2 (2 \pi z_2(s'))^2
\right),
\end{eqnarray}
where we have assumed $\dot{z}_1(s')>0$ and 
there is no singularities in $t$-plane
for deforming the contour $C_\pm$.
However, there are singularities
in the action where the $\cosh$ factor vanish.
We can see that 
if 
\begin{eqnarray}
 -\frac{1}{4} < {\rm Im} (z_2)- {\rm Re} (z_2) 
\frac{{\rm Im}(\dot{z}_1) }{{\rm Re} (\dot{z}_1)} < \frac{1}{4},
\label{cz}
\end{eqnarray}
there is no obstruction for the deformation of the contour.
If this is not the case, 
we can shift $z_2 \rightarrow z_2+ i n/2$,
where $n$ is an integer,
to satisfy the condition (\ref{cz}). 
Because the action is invariant under this,
we conclude that the free energy is 
\begin{eqnarray}
 f= 4 \pi N^{\frac{3}{2}} \int ds 
\left(
i k z_1(s) z_2(s) - i \zeta z_1(s) +2 \frac{1}{\dot{z}_1(s)} 
\left(
\frac{1}{16} +(z_2(s)+i h)^2
\right)
\right),
\label{freefunctional}
\end{eqnarray}
where $h \in {\mathbf Z}/2$ such that 
the condition
\begin{eqnarray}
 -\frac{1}{4} < {\rm Im} (z_2)- {\rm Re} (z_2) 
\frac{{\rm Im}(\dot{z}_1) }{{\rm Re} (\dot{z}_1)}+h < \frac{1}{4},
\label{cz2}
\end{eqnarray}
is satisfied.\footnote{ 
Note that 
\begin{eqnarray}
 {\rm Im} (z_2)- {\rm Re} (z_2) 
\frac{{\rm Im}(\dot{z}_1) }{{\rm Re} (\dot{z}_1)}+h = 
\frac{ {\rm Im} ((z_2+ih) \bar{\dot{z}}_1)  }{{\rm Re} (\dot{z}_1)}=
-k |\dot{z}_1|^2  \frac{ {\rm Re} (z_1) }{4 {\rm Re} (\dot{z}_1)}
.
\label{1/4boundrewrite}
\end{eqnarray}
Thus, if $z_2+ih \rightarrow \pm i/4$, then
$\frac{ {\rm Im} ((z_2+ih) \bar{\dot{z}}_1)  }{{\rm Re} (\dot{z}_1)} 
\rightarrow \pm1/4$, which is the edge of the bound \eqref{cz2}.
}

In the above derivation of the free energy $f$ \eqref{freefunctional}, the assumption that $\mathrm{Re}(z_1)$ is monotonically increasing (after the eigenvalues are rearranged so that the profile functions are piecewise continuous in $s$) is crucial.
This assumption is violated if the eigenvalue distribution has self-overlapping region after projected onto the real axis.
In this case \eqref{freefunctional} is corrected by the cross terms such
as $\log\sinh\pi(\lambda_i-\lambda_j)$ with $\lambda_i$ and $\lambda_j$
in two different segment with overlapping shades.

Here we will argue that such an overlapping configuration can not be 
the saddle point solution.
First suppose that the values of $\mathrm{Im}(z_1)$ are different for these two segments and denote the difference as $\mathrm{Im}(\Delta z_1)$.
We can evaluate the cross terms again using the formula \eqref{logcoshapprox1} and \eqref{logcoshapprox2}, but with the contour $C_\pm$ extended by a straight line $[\pm v(s_0),\pm v(s_0)+i\pi\sqrt{N}\mathrm{Im}(\Delta z_1(s_0))]$.
Since the integration of $\log (\cosh(t)e^{-t})$ over $\pi i$ vanishes, the contribution of $\mathrm{Im}(\Delta z_1)$ to the free energy depends on the remainder of $\sqrt{N}\mathrm{Im}(\Delta z_1)$ divided by $1$.
This implies that the profile functions obtained from the variation of the free energy depend non-trivially on the way to take the limit $N\rightarrow\infty$, hence the $N\rightarrow\infty$ will be ill defined.
To obtain a well defined large $N$ limit, we have to choose $\mathrm{Im}(\Delta z_1)=0$ at the level of the ansatz.
In this case, however, the original saddle point equation $\partial f/\partial(\lambda_i,{\widetilde\lambda}_i)$ will not be solved by the variational problem, as the degrees of freedom of the variations will be fewer than those for the smooth eigenvalue distributions for multiple segments.
The above argument shows that there are no solutions with overlapping
segments, at least, if we assume $f\sim N^{3/2}$.
Below we 
will consider only the cases without overlapping.

The saddle point equations are
\begin{eqnarray}
 0=i k z_2(s)-i \zeta  +2 \frac{\partial }{\partial s} 
\left(
\frac{1}{\dot{z}_1(s)^2} 
\left(
\frac{1}{16} +(z_2(s)+i h)^2
\right)
\right),
\end{eqnarray}
for the variation of $z_1$
with the following boundary condition:
\begin{eqnarray}
 0=\frac{1}{\dot{z}_1(s)^2} 
\left(
\frac{1}{16} +(z_2(s)+i h)^2
\right) \big{|}_{\rm boundary},
\end{eqnarray}
and
\begin{eqnarray}
 0=i k z_1(s) +4 
\frac{1}{\dot{z}_1(s)} 
(z_2(s)+i h),
\end{eqnarray}
for the variation of $z_2$,
which implies that
\begin{eqnarray}
z_2(s)+i h  =-i \frac{k}{4}  z_1(s) \dot{z}_1(s)
=-i \frac{k}{8}  \frac{\partial}{\partial s} (z_1(s)^2).
\end{eqnarray}
These implies that
\begin{eqnarray}
 0 &=& \frac{k^2}{8}  (z_1(s)^2)
-i (\zeta +i k h) (s-{s_0}) 
+2 
\frac{1}{\dot{z}_1(s)^2} 
\left(
\frac{1}{16} +(z_2(s)+i h)^2
\right) \nonumber \\
&=&
-i (\zeta +i k h) (s-{s_0})  
+
\frac{1}{8 \dot{z}_1(s)^2}, 
\end{eqnarray}
where ${s_0}$ is a complex integration constant.
Thus, we have
\begin{eqnarray}
z_1(s)= g \sqrt{s-{s_0}}+{z_0}, 
\,\,\,\,\,  \dot{z}_1(s)= g \frac{1}{2 \sqrt{s-{s_0}}}, 
\,\,\,\,\,  z_2(s)+ih= -i \frac{k g^2}{8}
- i {z_0} g \frac{k}{8 \sqrt{s-{s_0}}}, 
\end{eqnarray}
where ${z_0}$ is the integration constant and
\begin{eqnarray}
 g= \frac{1}{\sqrt{2i(\zeta+ikh)} }.
\end{eqnarray}
Note that because $z_1(s)$ should be a continuous function of $s$ 
we defined $\sqrt{s-s_0}$ as a continuous function of $s$
although we allowed the overall sign ambiguity.
This overall ambiguity should be fixed by the condition that
$z_1$ should be a monotonically increasing function of $s$.

To obtain the solutions, we need to specify the locations 
of the boundary points and the solutions 
should satisfy the condition (\ref{cz2}) everywhere.
Note that for general $\zeta$, above discussions are valid.
Indeed, the solutions for pure imaginary $\zeta$ also are included 
in the above solutions.

Now we assume $\zeta$ is real and there is only one segment
in the eigenvalue distributions. 
We will choose $s_0=i c$ where $c$ is real by shifting $s$.
Because there is one segment,
we choose the boundary points as $s=s_b$ and $s=s_b+1$.
Then, 
the boundary condition is 
\begin{eqnarray}
(z_2+i h)|_{s=s_b}= \gamma_1 \frac{i}{4},\quad
(z_2+i h)|_{s=s_b+1}=-\gamma_1\frac{i}{4},
\end{eqnarray}
where $(\gamma_1)^2=1$ representing a choice of 
the boundary values,\footnote{
The other possibility is 
$z_1(s)=g\sqrt{s}+z_0$ $(s=[0,1])$ which 
satisfies $\dot{z_1}(s=0)=\infty$
and $z_0$ is fixed by the boundary condition at $s=1$.
However, 
considering $s \sim 0$, we see that
for the condition (\ref{cz2})
${\rm Re } (z_0)=0$ is needed (see also \eqref{1/4boundrewrite}).
This is not satisfied for generic $\zeta/k$,
for example, with $h=0$, $z_0=0$ means $\zeta/k=1/4$.
}
which lead (assuming $\zeta \neq 0$) 
\begin{eqnarray}
&  k {z_0} \frac{1}{\sqrt{s_b-ic}} 
= - 2 \gamma_1 \frac{1}{g}- k g, \\
& k {z_0} \frac{1}{\sqrt{s_b+1-ic}} 
= 2 \gamma_1 \frac{1}{g}- k g.
\label{bd1}
\end{eqnarray}
We obtain $z_0$ from these boundary conditions:
\begin{eqnarray}
 1=\left( -\frac{1}{(\frac{2}{g}+\gamma_1 kg)^2}
+\frac{1}{(\frac{2}{g}-\gamma_1 kg)^2}
\right) (k z_0)^2
=\frac{8 \gamma_1 k}{(\frac{4}{g^2}- k^2 g^2)^2}(k z_0)^2,
\end{eqnarray}
which also lead 
\begin{eqnarray}
 s_b-ic=\gamma_1 \frac{1}{8k} \left( \frac{2}{g} - \gamma_1 kg \right)^2.
\end{eqnarray}
Thus, we find
\begin{eqnarray}
 s_b=-\frac{1}{2} -\gamma_1 \left(
h+\frac{1}{16} \frac{h}{m^2+h^2}
\right), \\
 c= \gamma_1 m \left(
-1+\frac{1}{16} \frac{1}{m^2+h^2}
\right).
\end{eqnarray}

Below, we will check that the solution 
is indeed a continuous function of $s$.
First, we define
\begin{align}
m\equiv& \frac{\zeta}{k},\\
 m_c \equiv& m +i h, \\
s' \equiv& \gamma_1 (s- s_b-\frac{1}{2}),
\end{align}
thus we find that
$s'=  -\gamma_1/2$ for $s=s_b$ and $s'= \gamma_1 /2$ for $s=s_b+1$.
With these, we find
\begin{eqnarray}
 z_1 = \frac{1}{\sqrt{-2 \gamma_1 k}} \left(
\sqrt{ \left( \frac{1}{16 m_c^2}-1 \right) +i  \frac{s'}{m_c} }
-4 \gamma_2 \left(  m_c + \frac{1}{16 m_c}   \right) 
\right),
\end{eqnarray}
and 
\begin{eqnarray}
  \dot{z}_1 =  \frac{i \gamma_1}{2  m_c \sqrt{-2\gamma_1 k}} 
\frac{1}{\sqrt{ \left( \frac{1}{16 m_c^2}-1 \right) + i  \frac{s'}{m_c} } },
\end{eqnarray}
which leads
\begin{eqnarray}
 z_2+ih= -\frac{1}{16 m_c}
+\gamma_2 \frac{1}{4}
\frac{1+\frac{1}{16 m_c^2}}{\sqrt{ \left( \frac{1}{16 m_c^2}-1 \right) +
i  \frac{s'}{m_c} }}.
\label{z2}
\end{eqnarray}
Here we introduced $\gamma_2$ which satisfies $(\gamma_2)^2=1$
for the sign ambiguity of $z_0$.
In order to satisfy the boundary condition $z_2 + ih =\pm i/4$,
we need 
\begin{eqnarray}
 \left. \sqrt{ \left( \frac{1}{16 m_c^2}-1 \right)  + i  \frac{s'}{
  m_c} } \right|_{s'=\mp \gamma_1/2}
=\gamma_2 \left( \frac{1}{4 m_c} \mp i \gamma_1 \right), 
\label{bc1}
\end{eqnarray}
at the boundaries.\footnote{
The condition is only for the sign because 
\begin{eqnarray}
 \left( \frac{1}{16 m_c^2}-1 \right)  -i \gamma  \frac{1}{2 m_c} 
=\left( \frac{1}{4 m_c}-i \gamma \right)^2, 
\end{eqnarray} 
for  $\gamma^2=1$.
}
This condition implies
$\gamma_2$ is fixed by the choice of the overall sign
in the l.h.s. of (\ref{bc1}).
Furthermore, we will see that for $m=\zeta/k >1/4$, these conditions are not consistent 
with the continuity of the factor $\sqrt{D}$ where 
\begin{eqnarray}
   D= \left( \frac{1}{16 m_c^2}-1 \right)  - i \frac{s'}{ m_c}, 
\end{eqnarray}
for $s'$.

As we will see below, ${\rm Re} (D)$ is negative for $m>1/4$.
Then, the phase $e^{i \theta} 
=\sqrt{D}/|\sqrt{D}|$ satisfies $\pi/4 < \theta < 3 \pi/4$
or $-\pi/4 > \theta > -3 \pi/4$ and we can easily see that
$|\sqrt{D}|>0$.
On the other hand, at the two boundaries,
we can see that $\mathrm{Im}(\sqrt{D})$ should have different signs
for $m>1/4$.
These are inconsistent with the continuity for $s'$.
For $h=0$, we easily see that ${\rm Re} (D)$ is indeed negative.
For $h \neq 0$,
we find 
\begin{eqnarray}
\label{eD}
 {\rm Re} (D) &=& \frac{1}{16 |m_c|^4} \left(
 ({\rm Re} (m_c))^2 -( {\rm Im} (m_c))^2
-16 |m_c|^4-16 |m_c|^2  {\rm Im} (m_c) s'
\right) \\
& \le& \frac{1}{16 |m_c|^4} \left(
 ({\rm Re} (m_c))^2 -( {\rm Im} (m_c))^2
-16 |m_c|^4+8 |m_c|^2  |{\rm Im} (m_c)| 
\right) <0,
\end{eqnarray}
where we have used $|s'| \le 1/2$ and 
$|{\rm Im} (m_c) |=|h| \ge 1/2$.
Therefore, there are no solutions 
for $m>1/4$.

We can also show that 
there are no solutions for
$m  \le 1/4$ and $h \neq 0$
because $ ({ \rm Re} (m_c))^2  -({ \rm Im} (m_c))^2 =m^2-h^2<0 $
and $ - |m_c|^4 +|m_c^2 h s' | < |m_c|^2 ( -(m^2+h^2) +|h/2 | )<0 $,
where we have used $|h| \ge 1/2$,
which implies $ {\rm Re} (D) <0$ using (\ref{eD}).
Therefore, only the possibility is for 
$m  \le 1/4$ and $h =0$.
For this case, 
we see that for $\gamma_1=-1$
${\rm Re} (\dot{z}_1)= 0$ at $s'=0$. Thus 
this solution violates the 
condition (\ref{cz2}) and we should set $\gamma_1=1$.
The solution is unique and given by 
\begin{eqnarray}
 z_1 = \frac{1}{\sqrt{-2 k}} \left(
\sqrt{ \left( \frac{1}{16 m^2}-1 \right) +i  \frac{s'}{m} }
-4 \gamma_2 \left(  m + \frac{1}{16 m}   \right) 
\right),
\label{z1}
\end{eqnarray}
where the sign ambiguity of the $\sqrt{ \left( \frac{1}{16 m^2}-1
\right) +i  \frac{s'}{m} }$ is 
fixed by requiring the condition ${\rm Re} (\dot{z}_1) \ge 0$
because 
we arranged the ordering of the eigenvalues such that $z_1(s)$ is
increasing function of $s$.

Finally, we will consider 
the multiple segments solutions.
The real part of such a solution should not intersect each other 
because of the extra interactions as explained before.
Then, the solutions are just a sum of the single segment solutions
with $N_a$ eigenvalues where $\sum_a N_a=N$.
However, the unique single segment solutions for $m<1/4$
with different $N$ always have an eigenvalue  such that ${\rm Re}
(\lambda)=0$.
Thus, there are no multiple segment solutions.\footnote{
So far, we have neglected a possibility that 
the solutions with different $h$ which have same
$z_1$ and $z_2$ at a boundary.
However, this is not possible 
because the cancellation of the boundary term
requires that $(z_2+ih)^2$ also should be same at the boundary.
}

Therefore, we conclude there is a unique solution 
for $m=\zeta/k < 1/4$, 
and no solutions for $m > 1/4$.
We can check that the solution for $m=\zeta/k < 1/4$ 
is indeed solution I 
in \cite{NST} which is derived also 
in appendix \ref{newvarprob}.
The free energy of this solution is
\begin{align}
f=\frac{\pi\sqrt{2k}}{3}N^{\frac{3}{2}} \Bigl(1+\frac{16\zeta^2}{k^2}\Bigr),
\label{FNto3/2}
\end{align}
as computed in \cite{NST}.
We can also evaluate the Wilson loop for the solution.
The exponent of Wilson-loop can be evaluated and is given as
\begin{align}
\label{wilsol1}
\langle W(C) \rangle\simeq & 
 e^{2\pi\sqrt{\frac{N}{k}}(\frac{1}{\sqrt{2}}+i\frac{2\sqrt{2}\zeta}{k})}.
\end{align}
Note that the real part of the exponent does not depend on $\zeta$
and for $\tilde{W}(C)$, the result is same.
We also note that
$\langle W(C^{-1}) \rangle =\langle W(C) \rangle$ where
$C^{-1}$ is the loop $C$ with the inverse direction.
This Wilson loop correspond to the BPS M2-brane wrapping the M-circle,
and $\sqrt{N}$ factor represents the tension of the M2-brane.

\section{'t Hooft limit}
\label{sec_tHooft}

In this section we consider the 't Hooft limit, $N,k,\zeta \rightarrow \infty$ with $N/k$ and $\zeta/k$ kept finite. 
Note that the mass of the chiral multiplets is proportional to $\zeta/k$, and hence finite in this limit.

\subsection{Strong 't Hooft coupling limit}

First we consider the strong 't Hooft coupling limit: $k\ll N$.
In this case it is easily seen that the eigenvalue distributions and the free energies reduce to those obtained for finite $k$ in section \ref{sec_finitek}.
Indeed, if we use the continuous notation $\lambda_i\rightarrow \lambda(s)$ with $s=i/N-1/2$ the saddle point equation \eqref{isaddle} is found to depend on $(N,k,\zeta)$ only through their ratio $(N/k,\zeta/k)$.
Hence, as the parameters in the strong 't Hooft coupling limit $1\ll (k,\zeta)\ll N$ can always be rescaled so that $1\ll N$ while $k$ and $\zeta$ are finite, we conclude that our analysis of the saddle point solutions and the free energies in the latter regime are still valid in the strong 't Hooft coupling limit.

\subsection{Weak 't Hooft coupling limit}
\label{sec_weak}
Second we consider the weak 't Hooft coupling limit: $k\gg N$.
In this limit, by assuming the balance between the first two terms and the second term in the saddle point equations \eqref{isaddle}, i.e. $(k\lambda_i-\zeta)\sim N\coth\pi(\lambda_i-\lambda_j)$, we find the following scaling behavior of $\lambda_i$
\begin{align}
\lambda_i=\frac{\zeta}{k}+{\cal O}\biggl(\sqrt{\frac{N}{k}}\biggr).
\end{align}
The explicit solution to the saddle point equations is given in the continuous notation as ($\lambda_i=x(s)+iy(s)$)
\begin{align}
x(s)&=\frac{\zeta}{k}+2\sqrt{\frac{N}{\pi k}}s,\nonumber \\
y(s)&=-2\sqrt{\frac{N}{\pi k}}s,
\label{sol_weak}
\end{align}
where $s\in(-1/2,1/2)$, together with the eigenvalue density $\rho(s)=(\frac{dx}{ds})^{-1}$ given with
\begin{align}
\rho(s)=\frac{8}{\pi}\sqrt{\frac{1}{4}-s^2}.
\label{rho_weak}
\end{align}
Below we first provide the derivation of this solution.
Then we evaluate the free energy and the vacuum expectation value of the Wilson loops on this solution.

To obtain the solution \eqref{sol_weak} and \eqref{rho_weak}, first let us shift the real/imaginary part of the eigenvalues as
\begin{align}
x_i=\frac{\zeta}{k}+u_i,
\end{align}
By assuming
\begin{align}
|u_i|,|y_i|\ll 1,
\label{smalldelta}
\end{align}
and expanding the trigonometric and hyperbolic functions up to ${\cal O}(u_i,y_i)$ we can simplify the saddle point equations \eqref{isaddle_rr} and \eqref{isaddle_ri} as 
\begin{align}
-y_i -\frac{1}{\pi k}\sum_{j=1(\neq i)}^N\frac{u_i-u_j}{(u_i-u_j)^2+(y_i-y_j)^2}&=0,\nonumber \\
u_i+\frac{1}{\pi k}\sum_{j=1(\neq i)}^N\frac{y_i-y_j}{(u_i-u_j)^2+(y_i-y_j)^2}&=0,
\end{align}
where we have neglected the deviations of ${\cal O}((N/k)^{3/2})$.
If we further pose the ansatz $y_i=-u_i$ and switch to the continuous notation
\begin{align}
u_i&\longrightarrow u\in I,\nonumber \\
\sum_{j=1(\neq i)}^N&\longrightarrow N\strokedint_I du\rho_u(u),\quad \Bigl(\int_Idu\rho_u(u)=1\Bigr),
\label{cont}
\end{align}
the saddle point equations reduce to the following single integration equation
\begin{align}
u=\frac{N}{2\pi k}\strokedint_Idu^\prime\rho_u(u^\prime)\frac{1}{u-u^\prime},
\end{align}
which is solved by
\begin{align}
I=(-\ell,\ell),\quad
\rho_u(u)=\frac{2k}{N}\sqrt{\ell^2-u^2}.
\end{align}
The real-positive parameter $\ell$ is determined from the normalization condition in \eqref{cont} as $\ell=\sqrt{\frac{N}{\pi k}}$.
We find that the weak coupling limit $k\gg N$ is indeed required for the consistency of the initial assumption \eqref{smalldelta}.
Changing the variable from $u$ to $s=\sqrt{\frac{\pi k}{N}}\frac{u}{2}$, we finally obtain the solution \eqref{sol_weak} with \eqref{rho_weak}.

The free energy \eqref{free_re} evaluated on this solution is
\begin{align}
f=N^2\Bigl[\log\frac{k}{4\pi N}+2\log \cosh\frac{2\pi\zeta}{k}+\frac{3}{2}+3\log 2\Bigr].
\end{align}
In the limit $\zeta\rightarrow 0$ the result coincide with that for the ABJM theory \cite{DMP1}.
We can also compute the Wilson loop as
\begin{align}
\langle W_\Box(C)\rangle&=e^{\frac{\zeta}{k}}\int^{\ell}_{-\ell}du \rho_{u}(u)e^{2\pi(u-iu)}\simeq e^{\frac{\zeta}{k}}\Bigl(1-\frac{i\pi N}{k}+{\cal O}\Bigl(\frac{N^2}{k^2}\Bigr)\Bigr),\nonumber \\
\langle {\widetilde W}_\Box(C)\rangle&\simeq e^{-\frac{\zeta}{k}}\Bigl(1+\frac{i\pi N}{k}+{\cal O}\Bigl(\frac{N^2}{k^2}\Bigr)\Bigr),
\end{align}
which are consistent with the results in \cite{KWY,DMP1} up to ${\cal O}(N/k)$ and the overall factor.

\section{Discussion}
\label{sec_disc}

In this paper we have studied the mass deformed ABJM theory in the large $N$ limit with various values of $(k,\zeta)$, using the saddle point approximation for the matrix model.
Let us rephrase our results, especially for finite $k$.

In section \ref{sec_largez} we have considered the limit $\zeta\gg k$.
In this parameter regime, since the mass of the matter multiplets is $m=\zeta/k$ we can integrate out these fields separately in the partition function.
As a result the saddle point equation gets extremely simplified, which is completely independent of $\zeta$.
Though the leading part of the free energy is fixed by the one-loop effects of the matter fields as $f\sim 4\pi\zeta N^2/k$, the eigenvalue distributions are still constrained by the saddle point equations.
We have also computed the vacuum expectation values of the Wilson loops in that saddle configuration, and found that they vanish due to non-trivial cancellation among the contributions from $N$ eigenvalues.

In the regime where both $\zeta$ and $k$ are finite, 
we found two different solutions.
One is the natural
extension of the above solution with $f\sim 4\pi\zeta N^2/k$
which exists for any $\zeta$ and $k$.
The other solution 
with $f\sim N^{3/2}$
which has the $AdS_4$ gravity dual exists only
for $\zeta/k < 1/4$ and 
coincides with the solution I in \cite{NST}.

Thus, the theory will be critical at $\zeta/k=1/4$
although the absolute value of Wilson loop does not depend on $\zeta/k$.
(As a matrix model, the eigenvalue distribution itself is 
the observable and becomes critical at the value.)
If we consider the large $N$ partition function on the solid torus \cite{SuTe}
which is obtained by cutting $S^3$,
we might see how the theory becomes critical $\zeta/k=1/4$
because the eigenvalues are fixed at the boundary of the 
solid torus. 
Of course, the analysis in the gravity dual is needed to 
understand the critical behavior.\footnote{
For $\zeta\in i\mathbb{R}$ the dual geometry which reproduces the free energy $F\sim N^{3/2}$ \eqref{FNto3/2} was studied in \cite{FP}, though the domain of validity of \eqref{FNto3/2} was not argued.
}
We hope to report on these in near future.

It is not clear that what is a correct solution for $\zeta/k > 1/4$.
One possibility is that it is the solution with $f  \sim N^2$ we found,
which implies that the free energy jumps between $\zeta/k< 1/4$
and $\zeta/k > 1/4$. For finite $N$, the partition function \eqref{ZN} will be continuous
with respect to $\zeta/k$, hence so is the free energy $f$.
However, 
this does not rule out the discontinuous change of the scaling exponent
of the large $N$ free energy $N^{3/2}\rightarrow N^2$
because the finite $N$ correction can make the free energy smooth.
Indeed, our solution which has the free energy of the order $N^{3/2}$
becomes singular at $\zeta=k/4$, thus it is not valid 
very near the point.
We expect that the analysis very near $\zeta=k/4$ including finite $N$ effects
gives a smooth free energy although we leave this problem for future work.
%

Other important property of the mass deformed ABJM theory 
is that it will describe the M2-M5 system.
Indeed, in the classical analysis \cite{GRVV}, the vacua are found to be 
given by a configuration which is 
a generalization of the fuzzy sphere
to a fuzzy $S^3$ which represents
the M5-brane \cite{T, GRVV}.
Thus, it would be natural to think the phase transition
at the critical value is due to the non-negligible effects of the 
spherical M5-branes and 
the compactified M5-branes would explain 
$f \sim N^2$ for $\zeta/k >1/4$.
We hope to report also on this in near future.

\section*{Acknowledgement}

We would like to thank Masazumi Honda, Seok Kim, Sanefumi Moriyama Shota
Nakayama and
Shuuichi Yokoyama for valuable discussions.

\appendix

\section{Evidence for another solution for $\zeta\gg k$}
\label{sec_anothersol}

Below we argue another possible way to solve the saddle point equations for $\zeta/k\gg 1$ \eqref{saddle_largez_r} and \eqref{saddle_largez_i},
\begin{align}
-kv_i-\sum_{\substack{j=1\\(j\neq i)}}^N\frac{\sinh 2\pi(u_i-u_j)}{\cosh 2\pi(u_i-u_j)-\cos 2\pi(v_i-v_j)}&=0,\label{real_appa} \\
ku_i+\sum_{\substack{j=1\\(j\neq i)}}^N\frac{\sin 2\pi(v_i-v_j)}{\cosh 2\pi(u_i-u_j)-\cos 2\pi(v_i-v_j)}&=0,\label{im_appa}
\end{align}
though the explicit expression is not found.

First we would like to assume $u_i<u_j$ for $i<j$, without loss of generality.
The key point is the following additional assumption: $v_i$ is large and varies more frequently than the real part $u_i$.
Under this assumption, we can compute the summation over $j$ in \eqref{real_appa} by approximate $u_i-u_j$ to be constant while $v_i-v_j$ spans a period of the cosine function, as
\begin{align}
\sum_{\substack{j=1\\(j\neq i)}}\frac{\sinh 2\pi(u_i-u_j)}{\cosh 2\pi(u_i-u_j)-\cos 2\pi(v_i-v_j)}&\approx \sum_{\substack{j=1\\(\neq i)}}\int_0^1dt\frac{\sinh 2\pi(u_i-u_j)}{\cosh 2\pi(u_i-u_j)-\cos 2\pi t}\nonumber \\
&=\sum_{\substack{j=1\\(j\neq i)}}\sgn(u_i-u_j)\nonumber \\
&=2i-1-N,
\end{align}
where in the second line we have used (the continuous version of) the formula \eqref{period1}, and in the third line we have used the fact $u_i<u_j\Leftrightarrow i<j$. 
Hence we can solve \eqref{real_appa} to obtain $v_i$
\begin{align}
v_i=-\frac{2N}{k}\Bigl(\frac{i}{N}-\frac{N+1}{2N}\Bigr)+\delta v\Bigl(\frac{i}{N}\Bigr).
\end{align}
Here $\delta v$ is some function of ${\cal O}(1)$.
If $\delta v(i/N)$ is randomly distributed and $k\ll N$, this $v_i$ indeed justifies the approximation for the summation above.

To determine the real part $u_i$ we have to solve the other equation \eqref{im_appa} (with the substitution of $v_i$)
\begin{align}
ku_i-\sum_{\substack{j=1\\(j\neq i)}}\frac{\sin(\frac{4\pi(i-j)}{k}-2\pi(\delta v(i/N)-\delta v(j/N)))}{\cosh 2\pi(u_i-u_j)-\cos(\frac{4\pi(i-j)}{k}-2\pi(\delta v(i/N)-\delta v(j/N)))}.
\end{align}
Though this equation is difficult to solve as it contains the random part $\delta v$,\footnote{
We cannot choose $\delta v=0$.
This fact is observed numerically, and also obvious at least for $k=1,2,4$; otherwise $u_i=0$ for all $i$ and contradict to the determination of $v_i$.
}
we have observed in the numerical analysis that the solution actually exist with several different-looking $\delta v$.

It is not clear whether the solutions of this type are relevant in the regime $\zeta/k\gg 1$.
For the numerical solutions we have obtained, however, we observe the following behavior of the free energy
\begin{align}
f-\frac{4\pi N^2\zeta}{k}\propto N^2
\end{align}
with some positive coefficient which is different for each solution.
Hence we conclude that the solution given in the section \ref{sec_largez} is more preferred in the saddle point approximation compared with these solution. 

Note that the solutions found in \cite{AR} is similar to this solution in the sense that the eigenvalue distribution is of ${\cal O} (N)$.

\section{Sub-leading part of solutions with $f\sim N^2$}
\label{O1_finitez}

In section \ref{sec_largez} and section \ref{sec_anyk} we have found the solutions of type
\begin{align}
\lambda_j\sim \frac{\zeta}{k}+i\Bigl(\frac{N}{k}+\frac{j}{N}\Bigr)+\cdots
\end{align}
(see \eqref{solI} and \eqref{solnew}), which manifestly solve the ${\cal O}(N)$ part of the saddle point equations.
In this appendix we show that these solution also solve the ${\cal O}(N^0)$ part of the saddle point equations, by explicitly determining the remaining part of the solution.
Hence we have a completely exact solution to the saddle point equations in the large $N$ limit.
Although they are irrelevant to the large $N$ analysis, the explicit solution would be helpful for the further analysis.

\subsection{Large $\zeta/k$}
\label{O1_finitez1}

Let us start with the simpler case, $\zeta/k\gg 1$, and determine the sub-leading profile of the saddle point solution $(g(s),n(j),\Delta)$ in \eqref{solI}.
The imaginary part of the saddle point equation is already of ${\cal O}(N^{-1})$, while the real part of the equations have the following terms of ${\cal O}(N^0)$
\begin{align}
-k\Bigl(\frac{i}{N}+\Delta+n(i)\Bigr)-\frac{2\pi}{N}\sum_{j=1(\neq i)}^N\frac{g(i/N)-g(j/N)}{1-\cos\frac{2\pi(i-j)}{N}}=0.
\end{align}
In the continuous notation
\begin{align}
&\frac{2\pi(i-N/2)}{N}\longrightarrow t\in(-\pi,\pi),\nonumber \\
&g\Bigl(\frac{i}{N}\Bigr)\longrightarrow g(t),\quad
n(i)\longrightarrow n(t)\in\mathbb{Z},\quad
\sum_{j=1(\neq i)}^N\longrightarrow \frac{N}{2\pi}\strokedint_{-\pi}^\pi dt^\prime,
\end{align}
the above equation is written as
\begin{align}
\strokedint_{-\pi}^\pi dt^\prime \frac{g(t)-g(t^\prime)}{1-\cos(t-t^\prime)}
=-\frac{k}{2\pi}\bigl(t+2\pi({\widetilde\Delta}+n(t))\bigr),
\label{largez_geq}
\end{align}
with ${\widetilde \Delta}=\Delta+1/2$.

To solve this equation, regard the last term in this equation as a linear transformation $P_1[\cdot]$ acting on a function
\begin{align}
P_1[g(t)]=\strokedint_{-\pi}^\pi dt^\prime\frac{g(t)-g(t^\prime)}{1-\cos(t-t^\prime)}.
\end{align}
We find the following series of the eigenfunctions and the eigenvalues of this operation
\begin{align}
P_1[\sin\alpha t]=2\pi \alpha \sin \alpha t.\quad\quad(\alpha\in\mathbb{N})
\end{align}
Assuming that $g(t)$ is expanded in these eigenfunctions and recalling the following identity used in \cite{NST}
\begin{align}
\sum_{\alpha=1}^\infty\frac{(-1)^{\alpha-1}}{\alpha}\sin \alpha t=\frac{t}{2},\quad\quad (-\pi<t<\pi)
\label{identityNST}
\end{align}
we obtain the following solution $({\widetilde \Delta},g(t))$ to the integration equation \eqref{largez_geq}
\begin{align}
{\widetilde\Delta}=0,\quad
g(t)=-\frac{k}{2\pi^2}\sum_{\alpha=1}^\infty\frac{(-1)^{\alpha-1}}{\alpha^2}\sin\alpha t=\frac{k}{2\pi^2}\mathrm{Im}\Li_2(-e^{it}).
\end{align}
With this choice the solution \eqref{solI} exactly solves the saddle point equations \eqref{saddle_largez_r} and \eqref{saddle_largez_i} in the large $N$ limit.

\subsection{Finite $\zeta/k$}
\label{O1_finitez2}

Next we consider the case with finite $\zeta/k$ \eqref{solnew} with the profile functions $(g(s),h(s),\Delta)$.
The strategy is the same as in appendix \ref{O1_finitez1}.
First we write down the four kinds of the summation in the saddle point equations \eqref{isaddle_rr} and \eqref{isaddle_ri} expanded with $g/N,h/N\ll 1$
\begin{align}
&\sum_{j(\neq i)}\frac{\sinh 2\pi(x_i-x_j)}{\cosh 2\pi(x_i-x_j)-\cos 2\pi(y_i-y_j)}
=\frac{2\pi}{N}\sum_{j(\neq i)}\frac{g_i-g_j}{1-\cos\frac{2\pi(i-j)}{N}}+{\cal O}\Bigl(\frac{1}{N}\Bigr),\nonumber \\
\nonumber \\
&\sum_j\frac{\sinh 2\pi(x_i+x_j)}{\cosh 2\pi(x_i+x_j)+\cos 2\pi(y_i-y_j)}
=\sum_j\frac{\sinh\frac{4\pi\zeta}{k}}{\cosh\frac{4\pi\zeta}{k}+\cos\frac{2\pi(i-j)}{N}}\nonumber \\
&\quad\quad\quad\quad +\frac{2\pi}{N}\sum_j\biggl[\frac{(g_i+g_j)\bigl(1+\cosh\frac{4\pi \zeta}{k}\cos\frac{2\pi(i-j)}{N}\bigr)}{\bigl(\cosh\frac{4\pi\zeta}{k}+\cos\frac{2\pi(i-j)}{N}\bigr)^2}
+\frac{(h_i-h_j)\sinh\frac{4\pi\zeta}{k}\sin\frac{2\pi(i-j)}{N}}{(\cosh\frac{4\pi\zeta}{k}+\cos\frac{2\pi(i-j)}{N})^2}\biggr]
+{\cal O}\Bigl(\frac{1}{N}\Bigr),\nonumber \\
\nonumber \\
&\sum_{j(\neq i)}\frac{\sin 2\pi(y_i-y_j)}{\cosh 2\pi(x_i-x_j)-\cos 2\pi(y_i-y_j)}
=\sum_{j(\neq i)}\frac{\sin\frac{2\pi(i-j)}{N}}{1-\cos\frac{2\pi(i-j)}{N}}-\frac{2\pi}{N}\sum_{j(\neq i)}\frac{h_i-h_j}{1-\cos\frac{2\pi(i-j)}{N}}+{\cal O}\Bigl(\frac{1}{N}\Bigr),\nonumber \\
\nonumber \\
&\sum_j\frac{\sin 2\pi(y_i-y_j)}{\cosh 2\pi(x_i+x_j)+\cos 2\pi(y_i-y_j)}
=\sum_j\frac{\sin\frac{2\pi(i-j)}{N}}{\cosh\frac{4\pi\zeta}{k}+\cos\frac{2\pi(i-j)}{N}}\nonumber \\
&\quad\quad\quad\quad+\frac{2\pi}{N}\sum_j\Bigl[-\frac{(g_i+g_j)\sinh\frac{4\pi\zeta}{k}\sin\frac{2\pi(i-j)}{N}}{\bigl(\cosh\frac{4\pi\zeta}{k}+\cos\frac{2\pi(i-j)}{N}\bigr)^2}
+\frac{(h_i-h_j)\bigl(1+\cosh\frac{4\pi\zeta}{k}\cos\frac{2\pi(i-j)}{N}\bigr)}{\bigl(\cosh\frac{4\pi\zeta}{k}+\cos\frac{2\pi(i-j)}{N}\bigr)^2}
\Bigr]
+{\cal O}\Bigl(\frac{1}{N}\Bigr),
\end{align}
where $g_i$ and $h_i$ are the abbreviations of $g(i/N)$ and $h(i/N)$ respectively.

After introducing the continuous notation replacing the discrete index $i$ and the summations
\begin{align}
\frac{2\pi(i-N/2)}{N}\rightarrow t\in(-\pi,\pi),\quad
\sum_{j=1(\neq i)}^N\longrightarrow \frac{N}{2\pi}\strokedint_{-\pi}^\pi dt^\prime,\quad
\sum_{j=1}^N\longrightarrow \frac{N}{2\pi}\int_{-\pi}^\pi dt,
\end{align}
the ${\cal O}(N^0)$ part of the saddle point equations can be written as
\begin{align}
&-\frac{kt}{2\pi}+k{\widetilde\Delta}
-\strokedint_I dt^\prime \frac{g(t)-g(t^\prime)}{1-\cos (t-t^\prime)}\nonumber \\
&\quad+\int_I dt^\prime \biggl[\frac{(g(t)+g(t^\prime))\bigl(1+\cosh\frac{4\pi \zeta}{k}\cos (t-t^\prime)\bigr)}{\bigl(\cosh\frac{4\pi\zeta}{k}+\cos (t-t^\prime)\bigr)^2}
+\frac{(h(t)-h(t^\prime))\sinh\frac{4\pi\zeta}{k}\sin (t-t^\prime)}{(\cosh\frac{4\pi\zeta}{k}+\cos (t-t^\prime))^2}\biggr]
=0,\nonumber \\
\nonumber \\
&-\strokedint dt^\prime\frac{h(t)-h(t^\prime)}{1-\cos (t-t^\prime)}\nonumber \\
&\quad+\int dt^\prime \Bigl[-\frac{(g(t)+g(t^\prime))\sinh\frac{4\pi\zeta}{k}\sin(t-t^\prime)}{\bigl(\cosh\frac{4\pi\zeta}{k}+\cos(t-t^\prime)\bigr)^2}
 +\frac{(h(t)-h(t^\prime))\bigl(1+\cosh\frac{4\pi\zeta}{k}\cos(t-t^\prime)\bigr)}{\bigl(\cosh\frac{4\pi\zeta}{k}+\cos(t-t^\prime)\bigr)^2}\Bigr]
=0,
\label{saddleint_finitez}
\end{align}
with ${\widetilde \Delta}=\Delta+1/2$.
To clarify the structure of the equations we introduce the following linear transformations
\begin{align}
P_1[g(t)]&\equiv \strokedint_I dt^\prime \frac{g(t)-g(t^\prime)}{1-\cos(t-t^\prime)},\nonumber \\
P_2[g(t)]&\equiv \int_I dt^\prime \frac{g(t)-g(t^\prime)}{\bigl(\cosh\frac{4\pi\zeta}{k}+\cos(t-t^\prime)\bigr)^2},\nonumber \\
P_3[g(t)]&\equiv \int_I dt^\prime \frac{\cos(t-t^\prime)}{\bigl(\cosh\frac{4\pi\zeta}{k}+\cos(t-t^\prime)\bigr)^2}(g(t)-g(t^\prime)),\nonumber\\
P_4[g(t)]&\equiv \int_I dt^\prime \frac{\sin(t-t^\prime)}{\bigl(\cosh\frac{4\pi\zeta}{k}+\cos(t-t^\prime)\bigr)^2}(g(t)-g(t^\prime)).
\end{align}
with which the saddle point equations \eqref{saddleint_finitez} are written compactly as
\begin{align}
\Bigl(-P_1-P_2-\cosh\frac{4\pi\zeta}{k}P_3\Bigr)[g(t)]+\sinh\frac{4\pi\zeta}{k} P_4[h(t)]&=\frac{kt}{2\pi}+k\Bigl(\frac{1}{2}-\Delta_2\Bigr)\nonumber \\
\sin\frac{4\pi\zeta}{k}P_4[g(t)]+\Bigl(-P_1+P_2+\cosh\frac{4\pi\zeta}{k}P_3\Bigr)[h(t)]&=0,
\end{align}
Since $e^{i\alpha t}$ $(\alpha\in\mathbb{Z})$ are eigenfunctions of these transformations
\begin{align}
P_a[e^{i\alpha t}]=\Lambda_{a,\alpha}e^{i\alpha t},
\end{align}
with $\Lambda_{a,\alpha}$ some constants, we shall pose the following ansatz
\begin{align}
g(t)=\sum_{\alpha\neq 0}A_\alpha e^{i\alpha t},\quad
h(t)=\sum_{\alpha\in\mathbb{Z}}B_\alpha e^{i\alpha t}.
\end{align}
Then, with the help of the identity for an infinite summation of the trigonometric functions \eqref{identityNST} we find that the saddle point equations are satisfied if the coefficients $A_\alpha$ and $B_\alpha$ satisfy the following equations
\begin{align}
\Bigl(-\Lambda_{1,\alpha}-\Lambda_{2,\alpha}-\cosh\frac{4\pi\zeta}{k}\Lambda_{3,\alpha}\Bigr)A_\alpha+\sinh\frac{4\pi\zeta}{k} \Lambda_{4,\alpha}B_\alpha&=\frac{k}{4\pi^2i}\frac{(-1)^{\alpha-1}}{\alpha},\quad (\alpha\neq 0)\nonumber \\
\sinh\frac{4\pi\zeta}{k}\Lambda_{4,0}B_0&=k\Bigl(\frac{1}{2}-\Delta_2\Bigr),\nonumber \\
\sin\frac{4\pi\zeta}{k}\Lambda_{4,\alpha}A_\alpha+\Bigl(-\Lambda_{1,\alpha}+\Lambda_{2,\alpha}+\cosh\frac{4\pi\zeta}{k}\Lambda_{3,\alpha}\Bigr)B_\alpha &=0.
\end{align}

\section{Solution for $\zeta/k < 1/4$
} 
\label{newvarprob}

In this appendix, we consider
the solutions with the free energy $f\sim N^{3/2}$ in a
similar way as in \cite{NST} 
in order to compare the results in this paper and the ones in \cite{NST}
easier. 
We use the continuous notation $\lambda_i\rightarrow \lambda(s)$ with $s\sim i/N-1/2\in(-1/2,1/2)$ and pose the following ansatz for the eigenvalue distribution
\begin{align}
\lambda(s)=\sqrt{N}x(s)+\Delta_e(s)+i(\sqrt{N}y_e(s)+y_o(s)),
\label{fullansatz}
\end{align}
with two odd functions $(x(s)$, $y_o(s))$ and two even functions $(\Delta_e(s), y_e(s))$ under $s\rightarrow -s$, all of which are of ${\cal O}(N^0)$.
We also assume $x(s)$ is a monotonically increasing function of $s$ without loss of generality due to the freedom of the re-numbering of the eigenvalues $\lambda_i\rightarrow \lambda_{\sigma(i)}$ with $\sigma$ any permutations.

Though the ansatz is a slight generalization of that in our previous work \cite{NST}, the process to determine the solution will look different.
Below we first substitute our ansatz to the free energy $f(\lambda,{\widetilde\lambda})$ \eqref{iff}.
The leading part of the free energy in the large $N$ limit can be regarded a functional of the profile functions $(x(s),\Delta(s),y_e(s),y_o(s))$.
Then we can obtain the new set of the ``saddle point equations'' from the variational problem of this functional.
Though the new procedure will be conceptually identical to the direct substitution of an ansatz to the original saddle point equations $\partial f/\partial \lambda_i$ and $\partial f/\partial{\widetilde \lambda}_i$ \eqref{isaddle}, we find that the derivation of the final set of the equations is substantially simplified.

As a result, we obtain a new boundary condition to the profile functions which were overlooked in the previous analysis and is essential to single out the solution.
We finally find that for $\zeta/k<1/4$ the solution is unique and coincide with the solution I in \cite{NST} and that for $\zeta/k>1/4$ there are no consistent solutions with our ansatz \eqref{fullansatz}.

With the substitution of the ansatz \eqref{fullansatz} the free energy is evaluated as
\begin{align}
f&=4\pi N^{3/2}H[x,\Delta_e,y_e,{\widetilde y}_o]+{\cal O}(N^{1/2})
\label{fH}
\end{align}
with
\begin{align}
&H[x,\Delta_e,y_e,{\widetilde y}_o]\nonumber \\
&\quad =\int ds\biggl[
\dot{x}^{-1}(2\Delta_e^2-kx\dot{y}_e\Delta_e)
+\zeta y_e+kh|x|-ky_e\Delta_e-kx{\widetilde y}_o
+\frac{\dot{x}}{\dot{x}^2+\dot{y}_e^2}\Bigl(\frac{1}{8}-2{\widetilde y}_o^2\Bigr)\biggr],
\label{H}
\end{align}
where the dot ``$\cdot$'' denotes the differential with respect to $s$.
We have also introduced the following abbreviation
\begin{align}
{\widetilde y}_o=y_o+h\sgn(s)-\dot{y}_e\Delta_e\dot{x}^{-1},
\end{align}
with $h\in\mathbb{Z}/2$ defined by $|{\widetilde y}_o|\le 1/4$.
We would like to note that the following integration identity is useful in the computation to derive the expression \eqref{fH} with \eqref{H}:
\begin{align}
\int_{s_0} ds \log\Bigl[\cosh\bigl(\sqrt{N}u(s)+v(s)\bigr) e^{-\sqrt{N}u(s)-v(s)}\Bigr]
\sim
\frac{1}{\sqrt{N}\dot{u}(s_0)}
\int_{-v(s_0)}^\infty dt \log\Bigl[\cosh (t) e^{-t}\Bigr],
\end{align}
for arbitrary complex functions $u(s),v(s)$ satisfying $\mathrm{Re}[u(s_0)]=0$ and $\mathrm{Re}[u(s)]>0$ for $s>s_0$.

Let us consider the extremization problem of the functional $H[x,\Delta_e,y_e,{\widetilde y}_o]$.
By differentiating with respect to the profile functions we obtain the following four differential equations
\begin{align}
\frac{d}{dx}\Bigl(\zeta y_e+\frac{1}{4}\frac{\rho}{1+y_e^{\prime 2}}\Bigr)+kh\sgn(x)&=0,\nonumber \\
kx\frac{d}{dx}(\rho\Delta_e)+\zeta\rho+\frac{d}{dx}\Bigl(\frac{y_e^\prime \rho^2}{(1+y_e^{\prime 2})^2}\Bigl(\frac{1}{4}-4{\widetilde y}_o^2\Bigr)\Bigr)&=0,\nonumber \\
-\frac{d}{dx}(kxy_e)+4\rho\Delta_e&=0,\nonumber \\
-kx-\frac{4\rho{\widetilde y}_o}{1+y_e^{\prime 2}}&=0.
\label{saddle_fH}
\end{align}
Here we have chosen $x$ as a fundamental variable rather than $s$, and introduced the eigenvalue density $\rho(x)=ds/dx$ in $x$ direction.
The differentials with respect to $x$ are abbreviated with primes ``$\prime$''.
In this notation we gain new degrees of freedom for the choice of the $x$-support $I_x$, as well as new constraints: $\rho(x)>0$ and the normalization condition
\begin{align}
\int_{I_x} dx\rho=1.
\label{normalizefH}
\end{align}
We also obtain the following constraints which come from the variation at the boundaries
\begin{align}
\rho\Delta_e\Bigr|_{\text{boundary}}=0,\quad\quad
{\widetilde y}_o\Bigr|_{\text{boundary}}=\pm \frac{1}{4}.
\label{bndryfH}
\end{align}
Interestingly our analysis (almost) derive the constraint $\Delta_e|_\text{boundary}=0$ which was posed just by hand in the previous analysis \cite{NST}.

The differential equations \eqref{fH} can be solved as follows.
From the first, third and fourth line of the equations we obtain
\begin{align}
\rho&=4(B-kh|x|-\zeta y_e)(1+y_e^{\prime 2}),\nonumber \\
\rho\Delta_e&=\frac{k}{4}\frac{d}{dx}(xy_e),\nonumber \\
\rho{\widetilde y}_o&=-\frac{kx(1+y_e^{\prime 2})}{4},
\end{align}
with $B$ an arbitrary constant.
Substituting these into the second line of \eqref{saddle_fH}, we obtain a differential equation containing only $y_e$, which is solved as
\begin{align}
y_e=\frac{B-kh|x|}{\zeta}-{\widetilde y}_e,
\end{align}
with
\begin{align}
{\widetilde y}_e&=\sqrt{\Bigl(1+\frac{k^2h^2}{\zeta^2}\Bigr)(x^2+2b|x|+a)}.
\end{align}
Here $a$ and $b$ are arbitrary real numbers.

\begin{figure}[ht!]
\begin{center}
\includegraphics[width=17cm]{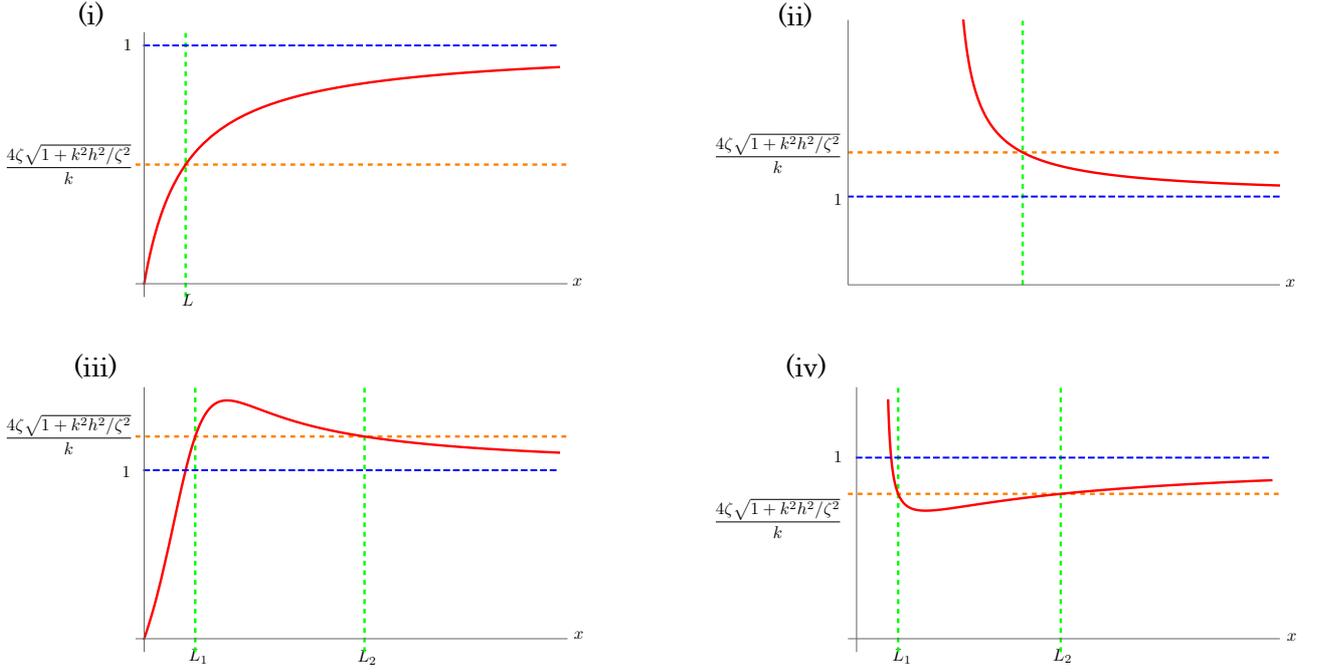}
\end{center}
\caption{
The behavior of the left-hand side of \eqref{yodbc} for (i) $a\ge 0$
and $b\ge 0$, (ii) $a<0$ and $b\le 0$, (iii) $a\ge 0$ and $b\le 0$ and
(iv) $a<0$ and $b\ge 0$.
}
\label{yodbcfig}
\end{figure}

Now we shall determine the moduli of the solution to the differential equations \eqref{saddle_fH}, which are $a,b,B\in\mathbb{R}$ together with the choice of the $x$-support $I_x$, from the normalization condition \eqref{normalizefH} and the boundary constraints \eqref{bndryfH}.
First we argue that the solution with any disconnected piece in $x\ge 0$ in $I_x$ is excluded from the boundary constraints \eqref{bndryfH}.
We focus on the second boundary condition ${\widetilde y}_o|_\text{boundary}$, which is explicitly written as 
\begin{align}
\frac{x}{\sqrt{x^2+2b|x|+a}}\biggr|_\text{boundary}=\frac{4\zeta\sqrt{1+\frac{k^2h^2}{\zeta^2}}}{k}.
\label{yodbc}
\end{align}
The behavior of the left-hand side as a function of $x$ is displayed in figure \ref{yodbcfig}.
For in the case (iii) and (iv), there exist two solutions $x=L_1,L_2$ for \eqref{yodbc} with $0\le L_1<L_2$.
Among them, the case (iii) ($a\ge 0,b\le 0$) is excluded as $|{\widetilde y}_o|>1/4$ for $L_1<x<L_2$ which contradicts to our initial assumption.
Hence the support $(L_1,L_2)$ could exist consistently only when the parameters satisfy $\zeta<k/4$, $h=0$, $a<0$ and $b>0$.
On the other hand, in the case of (iv) we can easily find that there are no solutions $(a,b,L_1,L_2)$ ($0\le L_1<L_2$) which also satisfy the first boundary condition in \eqref{bndryfH} $\rho\Delta_e(L_1)=\rho\Delta_e(L_2)=0$.
Hence we conclude that the $x$-support $I_x$ cannot have any disconnected segment in the region $x>0$; it must always be in the form of $I_x=(-L,L)$.

In the case $I_x=(-L,L)$, in addition to the boundary constraint at $x=L$, we also require the smoothness of the profile functions at $x=0$.
Indeed, any points where the profile functions are discontinuous require additional boundary constraints, with which the whole constraints become unsolvable as we have argued above.
Then it is obvious that the case $h\neq 0$ is excluded.
For the same reason we also find that $a$ and $b$ need to satisfy as $a>0$ and $b=0$.
Under these restrictions the moduli $(a,b,B,L_1)$ exist only when
\begin{align}
\zeta<\frac{k}{4},
\end{align}
(see plot (i) in figure \ref{yodbcfig}) and can be uniquely determined from the boundary constraint at $x=L$ and the normalization condition \eqref{normalizefH} as
\begin{align}
a=\frac{k}{32\zeta^2}\Bigl(1-\frac{16\zeta^2}{k^2}\Bigr),\quad
b=0,\quad
B=\frac{\sqrt{k}}{4\sqrt{2}}\Bigl(1+\frac{16\zeta^2}{k^2}\Bigr),\quad
I_x=\Bigl(-\frac{1}{\sqrt{2k}},\frac{1}{\sqrt{2k}}\Bigr).
\end{align}
with which the explicit expression for the profile functions are
\begin{align}
{\widetilde y}_e&=\sqrt{x^2+a},\label{ytildeC} \\
\rho&=4\zeta{\widetilde y}_e(1+{\widetilde y}_e^{\prime 2})=4\zeta\frac{d}{dx}(x{\widetilde y}_e),\label{rhoC} \\
 \Delta_{e}&= 
- \frac{1}{16m} \left(
1- \frac{B{\widetilde y}_e}{\zeta({\widetilde y}_e^2+x^2)}\right),\label{DeltaC} \\
 y_o &= - \frac{1}{16m} 
 \frac{B x}{\zeta({\widetilde y}_e^2+x^2)}.
\label{yoC}
\end{align}
The saddle point solution coincide with the solution I obtained in \cite{NST}.

The free energy of this solution is
\begin{align}
f=\frac{\pi\sqrt{2k}}{3}N^{\frac{3}{2}} \Bigl(1+\frac{16\zeta^2}{k^2}\Bigr),
\label{f1}
\end{align}
as computed in \cite{NST}.

We can check that the above solution of the saddle point equation corresponds to the solution we have introduced in section \ref{N^3/2gen}.
To see this we express $x$ as a function of $s$ by integrating $\rho=ds/dx$
\begin{align}
s(x)=\int_0^xdx\rho(x).
\end{align}
Using the explicit expression of $\rho$ \eqref{rhoC} we obtain
\begin{align}
 x =& \sgn(s) \sqrt{\frac{a}{2}} 
\sqrt{-1+\sqrt{1+ \frac{4 s^2 (2a+1)}{a^2}  }  },
\end{align}
where
\begin{align}
a=\frac{1}{2}\left(\frac{1}{16m^2}-1\right).
\end{align}
Hence ${\widetilde y}_e$ \eqref{ytildeC} is
\begin{align}
 \widetilde{y}_{e} =& \sqrt{\frac{a}{2}} 
\sqrt{1+\sqrt{1+ \frac{4 s^2 (2a+1)}{a^2}  }  }.
\end{align}
Now we can see that $x+iy_e$ coincides with $z_1(s)$ \eqref{z1}.
Similarly, $\Delta_e+iy_o$ (\eqref{DeltaC} and \eqref{yoC}) expressed in terms of $s$ coincide with $z_2(s)$ \eqref{z2}.

\end{document}